\documentclass[prb,twocolumn,superscriptaddress,citeautoscript,showpacs,amsart]{revtex4-2}

\usepackage{graphicx}
\usepackage{multirow}
\usepackage{physics}
\usepackage{color}
\usepackage{bm}
\usepackage{times}
\usepackage{amsmath,bm,amsfonts}
\usepackage{dcolumn}
\usepackage{graphicx}
\usepackage{latexsym}
\usepackage{cancel}
\usepackage{hyperref}

\usepackage{ulem} 
\usepackage{braket}

\def\ket#1{|#1\rangle }
\def\bra#1{\langle #1 |}
\def\n{\nonumber \\ }

\newcommand{\ma}{\sigma}
\newcommand{\V}{\vec}
\newcommand{\dt}{\delta}
\newcommand{\gm}{\gamma}
\newcommand{\ep}{\epsilon}

\newcommand{\f}{\frac}
\newcommand{\ta}{\theta}

\newcommand{\Dt}{\Delta}
\newcommand{\dg}{\dagger}

\newcommand{\A}{\alpha}
\newcommand{\B}{\beta}

\newcommand{\w}[1]{\omega_{#1}}
\newcommand{\al}[1]{\langle #1 \rangle}

\newcommand{\alg}[1]{\begin{align}#1\end{align}}

\newcommand{\nm}{\nonumber \\&}

\begin{document}

\title{Spin-phonon coupling and thermal Hall effect in the Kitaev model}

\author{Taekoo \surname{Oh}}
\email{taekoo.oh@ssu.ac.kr}
\affiliation{Department of Physics, Soongsil University, Seoul 06978, Korea}

\author{Naoto \surname{Nagaosa}}
\email{nagaosa@riken.jp}
\affiliation{RIKEN Center for Emergent Matter Science (CEMS), Wako, Saitama 351-0198, Japan}
\affiliation{Fundamental Quantum Science Program, TRIP Headquarters, RIKEN, Wako, Saitama, 351-0198, Japan}

\date{\today}

\begin{abstract}
The Kitaev model, which involves bond-direction-dependent spin interactions on a honeycomb lattice, has attracted significant interest due to its exact solvability and potential uses in quantum computing. A key feature of this model is the half-quantized thermal Hall conductivity (HQTHC) under a magnetic field perpendicular to the lattice; however, HQTHC only appears at low temperatures. Here, in the higher temperature range beyond the HQTHC regime, we theoretically suggest an extrinsic phonon contribution to the thermal Hall effect in the Kitaev model through skew-scattering of chiral phonons by scalar spin chirality, previously examined in Mott insulators. We demonstrate the emergence of scalar spin chirality from fluctuating spins, estimate the resulting field strength and its symmetric form applied to chiral phonons, and obtain the associated thermal Hall conductivity in semi-quantitative agreement with existing experiments. This work offers a fundamental understanding of how spin-phonon interactions influence strongly correlated systems.
\end{abstract}


\maketitle

\section{Introduction}

\begin{figure}[t]
    \centering
    \includegraphics[width=\linewidth]{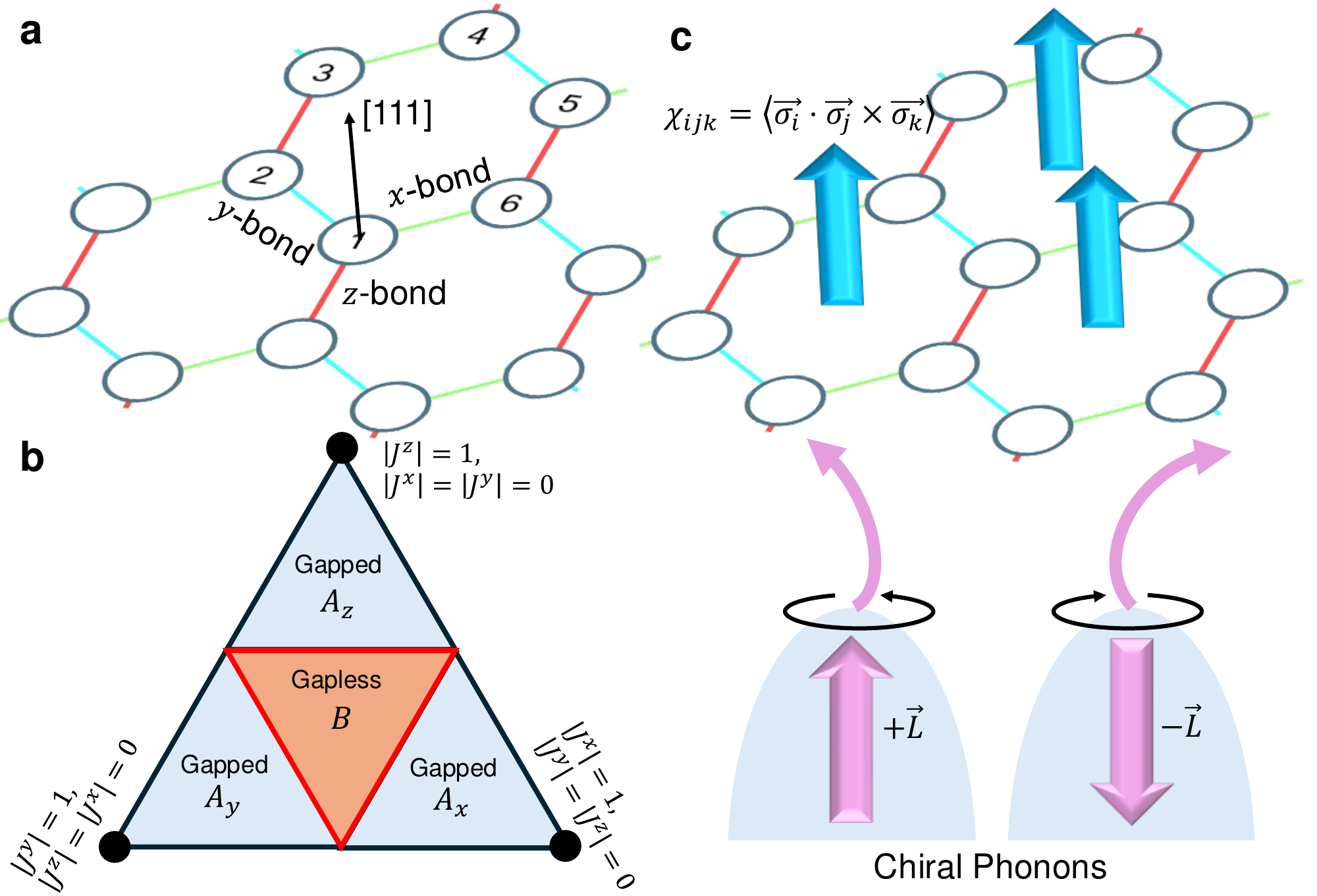}
    \caption{{\bf Phonon thermal Hall effect in the Kitaev model.}
    (a) The schematics of the Kitaev model. Green line is the $x$-bond, cyan line is the $y$-bond, and the red line is the $z$-bond. (b) The phase diagram of the Kitaev model. The orange lines are the phase boundary $|J^x| \leq |J^y| + |J^z|$, $|J^y| \leq |J^z| + |J^x|$, and $|J^z| \leq |J^x| + |J^y|$. Here, we concentrate on the gapless $B$ phase in a magnetic field. (c) The schematics of skew-scattering of chiral phonons by the scalar spin chirality.}
    \label{fig:1}
\end{figure}

The Kitaev model, characterized by spin exchange interactions that depend on the bond direction on a honeycomb lattice~\cite{kitaev2006anyons} [See Fig.~\ref{fig:1}(a)], has attracted significant attention over the past decades. 
The Kitaev model under a magnetic field is described by~\cite{zhang2022theory}
\alg{
H_0 = \sum_{\al{ij}} J_{ij}^{\A_{ij}} \ma_{i}^{\A_{ij}} \ma_j^{\A_{ij}} - \f{\vec{B}}{2}\cdot \sum_i \vec{\ma}_{i}. \label{eq:1}
}
Here, $\ma_i^{\A_{ij}}$ represents spin-$1/2$ operators with $\A_{ij} = x,y,z$, $J_{ij}^{\A_{ij}} = J^x, J^y, J^z$ denotes the interaction strength, and $\vec{B}$ is the applied magnetic field.
Without the magnetic field, the Hamiltonian is exactly solvable by fractionalizing the spins into Majorana fermions~\cite{kitaev2006anyons,feng2007topological,nasu2016fermionic}.
As shown in Fig.~\ref{fig:1}(b), when the interaction strengths satisfy $|J_x|\leq |J_y|+|J_z|$, $|J_y|\leq |J_z|+|J_x|$, and $|J_z|\leq |J_x|+|J_y|$, the Majorana fermion spectrum remains gapless ($B$ phase); otherwise, it becomes gapped ($A_x,A_y,A_z$ phases). 
When a magnetic field $\vec{B}\parallel[111]$ (perpendicular to the plane) is applied to the gapless $B$ phase, it opens a gap, leading to unique excitations and their non-Abelian statistics~\cite{kitaev2006anyons,takagi2019concept}. 
This phenomenon can be characterized by the emergence of edge modes and the associated half-quantized thermal Hall effect (HQTHE)~\cite{kitaev2006anyons,kane1997quantized,cappelli2002thermal}, and highlights the potential application of the Kitaev model in quantum computing. 
Consequently, the Kitaev model has been extensively considered in both 
theoretical~\cite{yao2007exact,kim2016crystal,modic2018chiral,gordon2019theory,gao2019thermal,metavitsiadis2020phonon,ye2020phonon,takahashi2021topological,zhang2021topological,hwang2022identification,yilmaz2022phase,li2022thermal,metavitsiadis2022optical,singh2023phonon,yu2023nematic,singh2024phonon,luo2024chiral,holdhusen2024emergent} and experimental~\cite{banerjee2016proximate,kasahara2018majorana,mcclarty2018topological,kasahara2018unusual,hentrich2018unusual,hentrich2019large,yamashita2020sample,sears2020ferromagnetic,czajka2021oscillations,wang2021evidence,yokoi2021half,suetsugu2022evidence,lefranccois2022evidence,kasahara2022quantized,bruin2022robustness,zhang2023sample,czajka2023planar,imamura2024majorana,hong2024phonon} contexts.

Among the promising candidates that realize the Kitaev model~\cite{trebst2022kitaev}, $\A$-RuCl$_3$ has been extensively investigated, particularly regarding thermal transport. 
Notably, the HQTHE has been reported in $\A$-RuCl$_3$ at $T \leq 6$ K under a magnetic field of $B \sim 8$ T~\cite{kasahara2018majorana,bruin2022robustness}.
Moreover, the thermal transport experiments have been conducted at temperatures up to $T \sim 80$ K~\cite{kasahara2018unusual,hentrich2018unusual,hentrich2019large,lefranccois2022evidence}, revealing remarkable features such as $\kappa_{yx}/T \sim 10^{-4}$ W$/$K$^2$m, a peak near $T=20\sim30$ K, and a long tail up to $T \approx 60 \sim 80 $ K.
However, several intriguing issues still remain: the sample dependence of HQTHE~\cite{yamashita2020sample,czajka2021oscillations,kasahara2022quantized,zhang2023sample}, the non-quantized thermal Hall conductivity (THC) attributed to topological magnons~\cite{zhang2021topological}, and the large THC from Kitaev-Heisenberg paramagnons~\cite{hentrich2018unusual,hentrich2019large} and phonons~\cite{lefranccois2022evidence}.
Especially, the phonon thermal Hall effect must be considered due to its distinct temperature dependence. 
However, a magnetic ordering is necessary so far~\cite{li2022thermal}, so the phonon thermal Hall effect in the high-temperature regime remains unclear.
Further investigations into the phonon thermal Hall effect within the Kitaev model are necessary. 

In this context, we propose an extrinsic origin of the phonon thermal Hall effect in the Kitaev model for temperatures above $10$ K, beyond the HQTHE regime. 
This mechanism, known as skew-scattering of chiral phonons by the scalar spin chirality (SSC), is illustrated in Fig.~\ref{fig:1}(c) and has been previously studied in multiferroic Mott insulators like YMnO$_3$~\cite{kim2024thermal,oh2024phonon}. 
Here, we extend the mechanism into the high-temperature range of the Kitaev model.

The SSC $\chi_{ijk}=\al{\vec{\ma}_i\cdot\vec{\ma}_j\times\vec{\ma}_k}$ is a composite order parameter formed by three different spins at sites $i$, $j$, and $k$. 
It distinguishes the noncoplanar spin structure~\cite{wen1989chiral,kawamura1992chiral,shindou2001orbital,taguchi2001spin,lee2006doping,kawamura2010chirality,nagaosa2012gauge,nagaosa2012emergent}. 
Although each spin individually is not ordered, the SSC can be finite only by the spin fluctuations.
Since SSC breaks the time-reversal symmetry, it can induce transport phenomena such as anomalous Hall and Nernst effects~\cite{bruno2004topological,neubauer2009topological,nagaosa2010anomalous,kanazawa2011large,ishizuka2018spin,ishizuka2021large}.
Under an applied magnetic field, spin fluctuations within each hexagonal plaquette of the lattice can generate a finite SSC. 
Specifically, in both ground and excited states of Kitaev model, a magnetic field induces a finite SSC.
Because the electronic many-body wavefunction in each plaquette deforms due to SSC, a Berry connection emerges, producing an emergent field that directly couples to chiral phonons. 
This spin-phonon coupling, known as a Raman-type interaction~\cite{agarwalla2011phonon,saito2019berry}, scatters chiral phonons with opposite angular momentum asymmetrically, leading to a sizable thermal Hall effect comparable to that generated by spins and magnons~\cite{katsura2010theory,onose2010observation}. 
In an ideal scenario, the emergent field can reach $200$ T, and $|\kappa_{yx}|$ can rise to $10^{-2}$ W/Km~\cite{oh2024phonon}.
Previously, spin-phonon coupling in the Kitaev model~\cite{metavitsiadis2020phonon,li2022thermal,metavitsiadis2022optical,singh2023phonon,singh2024phonon} was mainly considered to either renormalize the phonon spectrum or  analyze phonon dynamics. 
When spin-phonon coupling was included in the context of the thermal Hall effect, the primary mechanism was regarded as the  intrinsic contribution from the Berry curvature of magnon and phonon bands. 
However, we highlight that in our work, spin-phonon coupling arises from the finite SSC driven by spin fluctuations under a magnetic field, leading to an extrinsic skew-scattering of the chiral phonons.
This mechanism can be relevant to a wide range of Kitaev model candidates.

\begin{figure*}[ht]
    \centering
    \includegraphics[width=\linewidth]{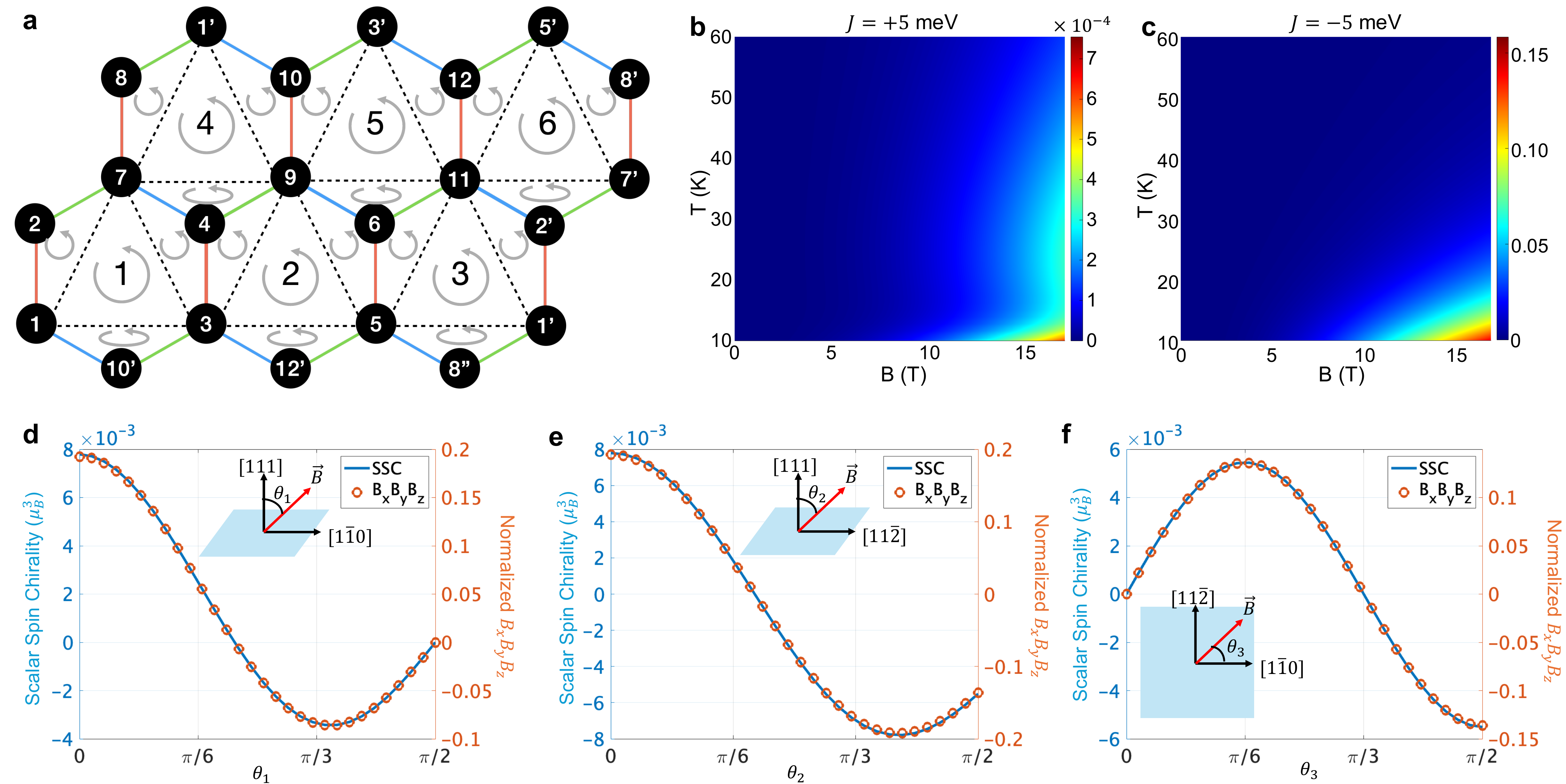}
    \caption{{\bf The scalar spin chirality in the Kitaev model.}
    (a) The schematics of a $2 \times 3$ unit cell (12 spins) in the periodic boundary condition. We divide each hexagon into four triangles and compute the SSC at each triangle. The direction of $(abc)$ follows the gray arrows. We add up all SSC and acquire the SSC per hexagon $\chi_p$.
    (b-c) The SSC per hexagon $\chi_p$ computed by the exact diagonalization as a function of magnetic field ($B$) and temperature ($T$), for (b) $J=+5$ meV (antiferromagnetic) and (c) $J=-5$ meV (ferromagnetic). (d-f) The SSC with different magnetic field directions at $J=-5$ meV, $T=20$ K, and $B = 8$ T. The blue lines are the SSC and the orange points are the normalized $B_xB_yB_z$. The field direction is in the plane including (d) $[111]$ and $[1\bar10]$, (e) $[111]$ and $[11\bar2]$, and (f) $[1\bar10]$ and $[11\bar2]$, as shown in their insets.
    }
    \label{fig:2}
\end{figure*}

The rest of the paper is structured as follows.
In Sec.~\ref{sec:2}, we employ the exact diagonalization of Eq.~(\ref{eq:1}) to compute the temperature and magnetic field dependence of SSC in the Kitaev model. 
In Sec.~\ref{sec:3}, we estimate the emergent field strength, and in Sec.~\ref{sec:4}, we examine its symmetric form to establish the Raman-type interaction. 
In Sec.~\ref{sec:5}, we incorporate the Raman-type interaction into the semiclassical transport theory to compute the THC. 
Our results highlight key features such as $\kappa_{yx}/T \sim 10^{-4}$ W$/$K$^2$m, a peak near $T=15$ K, and a long tail extending up to $40$ K, which are in the semi-quantitative agreement in the order of magnitude with the experimental observations.
Finally, we discuss our findings and present our conclusion in Sec.~\ref{sec:6}.

\section{\label{sec:2} scalar spin chirality in the Kitaev model}

First, we examine how temperature and magnetic field affect the SSC in the Kitaev model in Eq.~(\ref{eq:1}).
It is important to note that the bond-direction-dependent interaction (Kitaev interaction) is the key ingredient of SSC, but the usual superexchange is not.
Furthermore, while previous theoretical studies have computed SSC in the ground state of the Kitaev models~\cite{modic2018chiral,holdhusen2024emergent,luo2024chiral}, here we focus on the effects at finite temperatures.

We consider a system of $12$ spins arrayed in a $2\times3$ unit cell honeycomb lattice, as shown in Fig.~\ref{fig:2}(a), and compute the full spectrum $\{E_i\}$ along with the corresponding eigenvectors $\{\ket{\psi_i}\}$ using exact diagonalization.  
From these, we construct the density matrix $\rho = \sum_i \f{e^{-\B E_i}}{Z}\ket{\psi_i}\bra{\psi_i}$ with $Z = \sum_j e^{-\B E_j}$. 
Then, we compute the SSC per plaquette $\chi_p = \sum_{(abc)}\Tr(\rho\vec{\ma_a}\cdot \vec{\ma_b}\times\vec{\ma_c})/6$, where the summation runs over the triangular plaquettes $(abc)$ shown in Fig.~\ref{fig:2}(a). 
We set the coupling strength $J^x=J^y=J^z=J=\pm 5$ meV, with temperature range $10$ K $<T< 60$ K, and magnetic field $0$ T $<B<17$ T.

The results for $\vec{B} \parallel [111]$ are shown in Figs.~\ref{fig:2}(b) and \ref{fig:2}(c), revealing four main observations.
First, the SSC remains positive throughout all regions. 
Second, the SSC is one or two orders larger in the ferromagnetic Kitaev model with $J=-5$ meV, compared to the antiferromagnetic case with $J=5$ meV.
Next, the SSC increases as the magnetic field strengthens because the magnetic field breaks the time-reversal symmetry.
Lastly, SSC decreases with rising temperature because excited states with opposite scalar spin chiralities are populated more evenly at high temperature.

Because many Kitaev candidates are considered ferromagnetic~\cite{kim2022alpha}, we will now only use the ferromagnetic Kitaev model.

The results for different magnetic field directions at $J=-5$ meV, $T=20$ K and $B=8$ T are shown in Figs.~\ref{fig:2}(d-f). 
When we define $a=[1\bar10]$, $b=[11\bar2]$, and $c=[111]$, Fig.~\ref{fig:2}(d) shows the $ac$-plane field, Fig.~\ref{fig:2}(e) shows the $bc$-plane field, and Fig.~\ref{fig:2}(f) shows the $ab$-plane field results, as depicted in their insets.
Here, the $a$-axis is parallel to one of the bonds~\cite{hwang2022identification,imamura2024majorana}.
The blue lines represent SSC, the orange points show the normalized $B_xB_yB_z$. 
When $\vec{B} = B(\sin\ta\cos\phi,\sin\ta\sin\phi,\cos\ta)$, the normalized $B_xB_yB_z$ is $\sin^2\ta\cos\ta\sin\phi\cos\phi$.

We note two important points. 
First, the angular dependence in Fig.~\ref{fig:2}(d) is $\cos\ta_1(-1+5\cos(2\ta_1))$, in Fig.~\ref{fig:2}(e) it is $2\cos^3\ta_2 -3 \sin^2\ta_2\cos\ta_2 -\sqrt2 \sin^3\ta_2$, and in Fig.~\ref{fig:2}(f) it is $\sin3\ta_3$.
This matches the angular dependence of $B_xB_yB_z$. 
Thus, $\chi\propto B_xB_yB_z \propto \Dt$, where $\Dt$ is the gap of the ground state of the Kitaev model.
Next, as a corollary, a purely in-plane magnetic field can also induce the SSC, which can lead to the planar thermal Hall effect.
We anticipate the planar thermal Hall effect has angular dependence $\sin3\ta_3$ as shown in Fig.~\ref{fig:2}(f), consistent with previous experiments~\cite{imamura2024majorana}.

\section{\label{sec:3} The emergent field strength from the scalar spin chirality}

As SSC appears in the Kitaev model, it leads to an associated emergent field. 
In this section, we estimate the strength of the emergent field. 
Before proceeding, we briefly review the physics of emergent fields and the Raman-type interaction.

The complete Hamiltonian of a solid is described by $H = H_{\text{nu}} + H_{\text{el}} + V$. 
Here, $H_{\text{nu}} = -\sum_a \nabla_{R_a}^2/2M_a$ represents the nucleic kinetic energy, $H_{\text{el}} = -\sum_i \nabla_{r_i}^2/2m_e$ describes the electronic kinetic energy, and $V = -\sum_{i,a}Z_a/r_{ia} + \sum_{i>j}1/r_{ij} + \sum_{a>b}Z_aZ_b/r_{ab}$ accounts for the Coulomb interactions. 
$M_a$ and $Z_a$ denote the mass and charge of the nuclei, respectively, while $m_e$ denotes the electron mass. The sets $\{\V r_i\}$ and $\{\V R_a\}$ correspond to the positions of electrons and nuclei, respectively.
Since electron energy scales are much larger than those of phonons in solids, the Born-Oppenheimer approximation~\cite{oppenheimer1927quantentheorie} can be applied. 
In this approximation, the wavefunction of $H$ factorizes into electronic and nuclear wavefunctions: $\ket{\Psi} \approx \ket{\psi_{\text{el}}(\{\vec{r}_i\}, \{\vec{R}_a\})} \ket{\psi_{\text{nu}}(\{\vec{R}_a\})}$.
By integrating out the electronic wavefunction, an effective Hamiltonian for the nuclear motion is obtained.

The Raman-type interaction occurs when corrections to the Born-Oppenheimer approximation are considered. Specifically, after integrating out the electronic wavefunction, the Berry connection, $\vec{a}_a = -i\bra{\psi_{\text{el}}}\nabla_{R_a}\ket{\psi_{\text{el}}}$, must be retained. 
$\vec{a}_a$ was initially ignored because the characteristic length scales of electrons $l_{\text{el}}$ and nuclei $l_{\text{nu}}$ were thought to be vastly different, which implies $\grad_{R_a} \ket{\psi_{\text{el}}} \approx 0$.
However, it turns out that the Berry connection becomes significant because the length scales are not extremely different, with their ratio given by $l_{\text{el}}/l_{\text{nu}} = (M_a/m_e)^{1/4} \lesssim 10$.
Accordingly, the corrected Born-Oppenheimer approximation takes the following form under an external magnetic field~\cite{mead1979determination,mead1992geometric,zhang2010topological,qin2012berry,saito2019berry}:
\alg{
H_{BO} = \sum_{a} \f{(\vec{P}_a-Z_a\vec{A}_a+\vec{a}_a)^2}{2M_a} + E_{\text{el}}(\{\vec{R}_a\}).
}
Here, $\vec{A}_a$ denotes the external vector potential, $E_{el}(\{\vec{R}_a\})$ represents the electronic energy for fixed $\{\vec{R}_a\}$.
In the single atom, the Berry connection $\vec{a}_a$ completely cancels the external vector potential $\vec{A}_a$, leading to the screening effect of electrons around the nucleus.
In the lattice, however, the cancellation is incomplete, leaving the residual Raman-type interaction of the form $\vec{P}_a \cdot \vec{a}_a'$, where $\vec{a}_a' = Z_a\vec{A}_a - \vec{a}_a$. 
The Raman-type interaction generates an emergent field $\vec{B}_e = \grad\times \vec{a}_a'$, which plays a key role in phonon transport.
Notably, $\vec{a}_a$ does not appear if either inversion or time-reversal symmetry is preserved.

\begin{figure}[t]
    \centering
    \includegraphics[width=\linewidth]{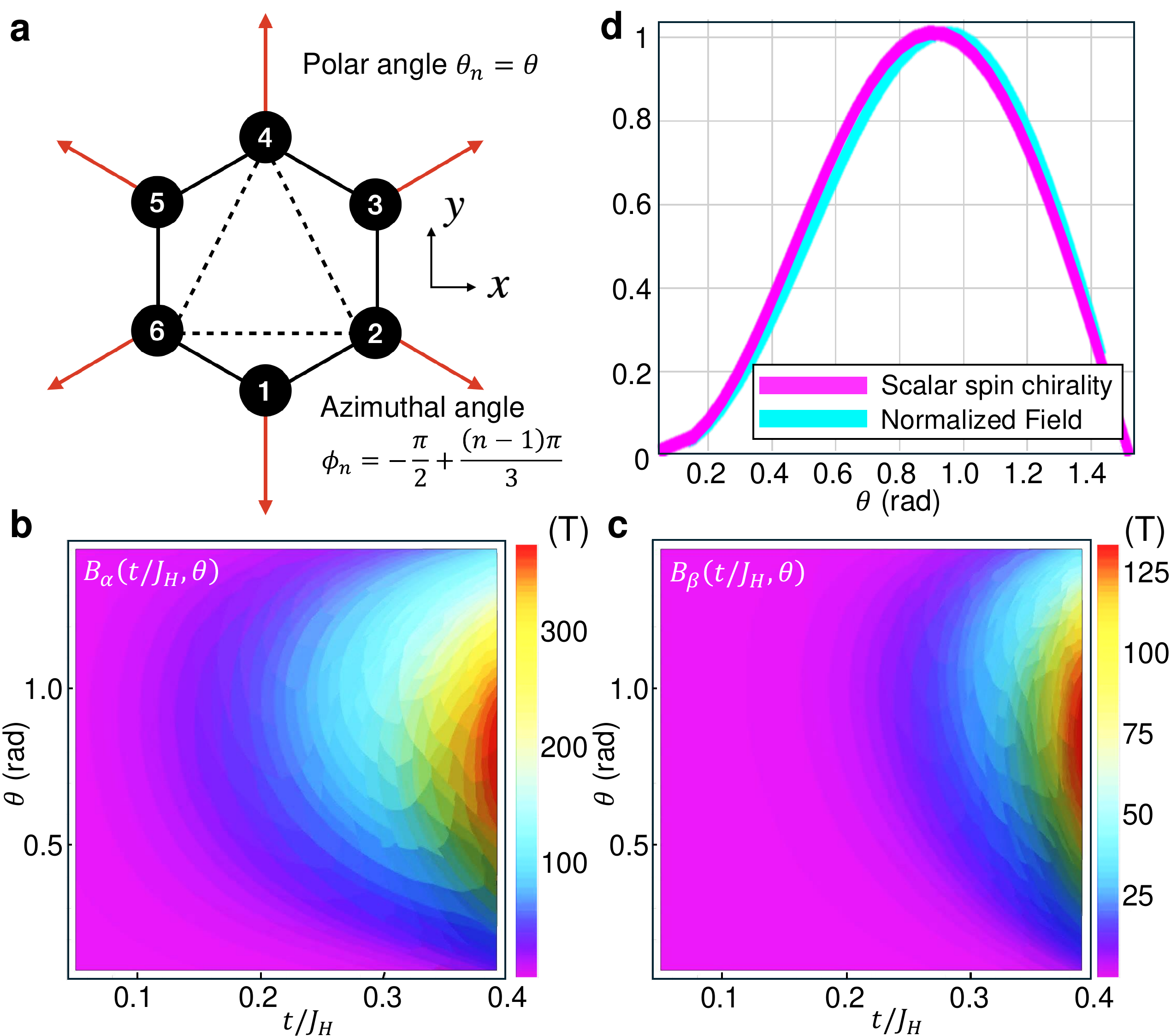}
    \caption{{\bf The estimation of emergent field strength.}
    (a) The schematics of spin configuration in double exchange model. For $n$th spin, the polar angle is $\ta_n = \ta$ and the azimuthal angle is $\phi_n=-\pi/2+(n-1)\pi/3$. (b-c) The emergent field strength from $B_\A(t/J_H,\theta)$ and $B_\B(t/J_H,\ta)$ in Eq.~(\ref{eq:B}) in the unit of Tesla. 
    (d) The comparison between $\chi$ and normalized field strength at $t/J_H = 0.23$ as a function of the polar angle $\ta$.
    }
    \label{fig:3}
\end{figure}

Using the established framework, we now estimate the emergent field strength. Here, we assume that the SSC from both quantum and classical spins can contribute equally to the emergent field and Raman-type interaction. 
This is because the key ingredient of the extrinsic phonon thermal Hall effect is not the spin ordering but the scalar spin chirality. 
Since the SSC can emerge from either spin fluctuations or spin ordering, we can estimate the emergent field strength based on spin-ordered states. Therefore,
to model the insulating state, we employ the double exchange model with classical spins:
\alg{
H_1 = \sum_{ij} t_{ij}c_{i\A}^\dg c_{j\A} - J_H \sum_{i} \vec{S}_i \cdot c_{i\A}^\dg \vec{\ma}_{i,\A\B} c_{j\B}.
}
Here, $t_{ij}$ is the hopping integral between site $i$ and $j$, $J_H$ is the double exchange strength, $c_{i\A}$ is the electron operators, $\vec{S}_i$ is the classical spins, and $\vec{\ma}_i$ is the electron spins. The hopping integral comprises two parts,
$t_{ij} = t+\dt t_{ij}$, where $t$ is the hopping integral at the equilibrium nuclear position, and  $\dt t_{ij}$ accounts for the variation of the hopping integral caused by the nuclei displacement.
Here, we assume that $\dt t_{ij}$ depends solely on the nuclear position $\{\vec{R_a}\}$.
The spin configuration is depicted in Fig.~\ref{fig:3}(a), where the $n$th spin in the hexagon has a polar angle $\ta_n = \ta$ and an azimuthal angle $\phi_n = -\pi/2 + (n-1)\pi/3$. 
In this configuration, the total SSC of the hexagon is $\chi =3\sqrt3 \cos\ta\sin^2\ta$.
Since the system is insulating, we assume six electrons, corresponding to half-filling. 
The control parameters are chosen as $t/J_H \in (0,0.4]$ and $\ta \in (0,\pi/2)$.
Based on previous studies, we take $J_H \approx 0.35$ eV, which is estimated from the Mott gap about $U\approx 1.9$ eV~\cite{rojas1983hall,plumb2014alpha,sandilands2016spin,zhou2016angle,sinn2016electronic}.

To analyze the emergent field strength, we first identify the second-order perturbed single-particle eigenfunctions $\ket{\varphi_\gm} \equiv \ket{\varphi_\gm(t/J_H,\ta,\dt t_{ij})}$ ($\gm=1,...,12$) of $H_1$, treating $\dt t_{ij}$ as a perturbation. 
The many-body wavefunction is then constructed as a Slater determinant of the six lowest-energy single-body states, $\ket{\psi_{\text{el}}} = \f{1}{6!}\ep_{\gm_1\gm_2\gm_3\gm_4\gm_5\gm_6}\ket{\varphi_{\gm_1}}\ket{\varphi_{\gm_2}}\ket{\varphi_{\gm_3}}\ket{\varphi_{\gm_4}}\ket{\varphi_{\gm_5}}\ket{\varphi_{\gm_6}}$.
The Berry connection for the $a$th nucleus ($a=1,...,6$) is then given by
\alg{
\vec{a}_a =& \sum_{(ijklmn)} (\grad_{R_a} \dt t_{ij})[\A(t/J_H,\ta)(\dt t_{jk}-\dt t_{ni}) \nm + \B(t/J_H,\ta)(\dt t_{mn}-\dt t_{kl})], \label{eq:A}
}
where $(ijklmn) = (123456)$ and its cyclic permutations. Approximating $\dt t_{ij} \approx \grad_{R_b} \dt t_{ij} \cdot \vec{u}_b$ with the displacement $\vec{u}_b$, the emergent field strength $B_e \approx B_\A + B_\B$ is estimated as
\alg{
B_{\A}(t/J_H,\ta) \approx& (\grad_{R_a} \dt t_{ij})^2 \A(t/J_H,\ta),\n 
B_\B (t/J_H,\ta)\approx& (\grad_{R_a} \dt t_{ij})^2 \B(t/J_H,\ta). \label{eq:B}
}
Taking $|\grad_{R_a} \dt t_{ij}| \approx 100$ meV/$\AA$~\cite{coropceanu2007charge}, we numerically compute $B_\A$ and $B_\B$ as functions of $t/J_H$ and $\ta$ in Figs.~\ref{fig:3}(b) and 3(c), expressed in Tesla. 

The emergent field strength increases with $t/J_H$ and peaks near $\ta \approx 1$ rad when $t/J_H \leq 0.3$. 
Notably, the emergent field strength can reach several hundred Tesla. 
To further analyze the correlation between the emergent field and SSC, we compare the emergent field strength at $t/J_H = 0.23$ with the SSC in Fig.~\ref{fig:3}(d). 
Both quantities peak near $\ta \approx 1$ and vanish at $\ta = 0$ and $\ta = \pi/2$, emphasizing their intrinsic correlation~\cite{oh2024phonon}.

\section{\label{sec:4} The symmetric form of emergent field from the scalar spin chirality}

Despite the emergence of SSC, the emergent field may disappear due to symmetry. 
Especially, Eq.~(\ref{eq:A}) becomes zero when $\dt t_{jk} = \dt t_{ni}$ and $\dt t_{mn} = \dt t_{kl}$. 
The condition is naturally satisfied when nuclei displacements preserve the threefold rotation symmetry $3_z$ about the $z$ axis, rendering $\vec{a}_a$ vanish for all $a$. 
Thus, symmetry analysis is essential for determining the allowed symmetric forms of the emergent field.

\begin{figure}[t]
    \centering
    \includegraphics[width=\linewidth]{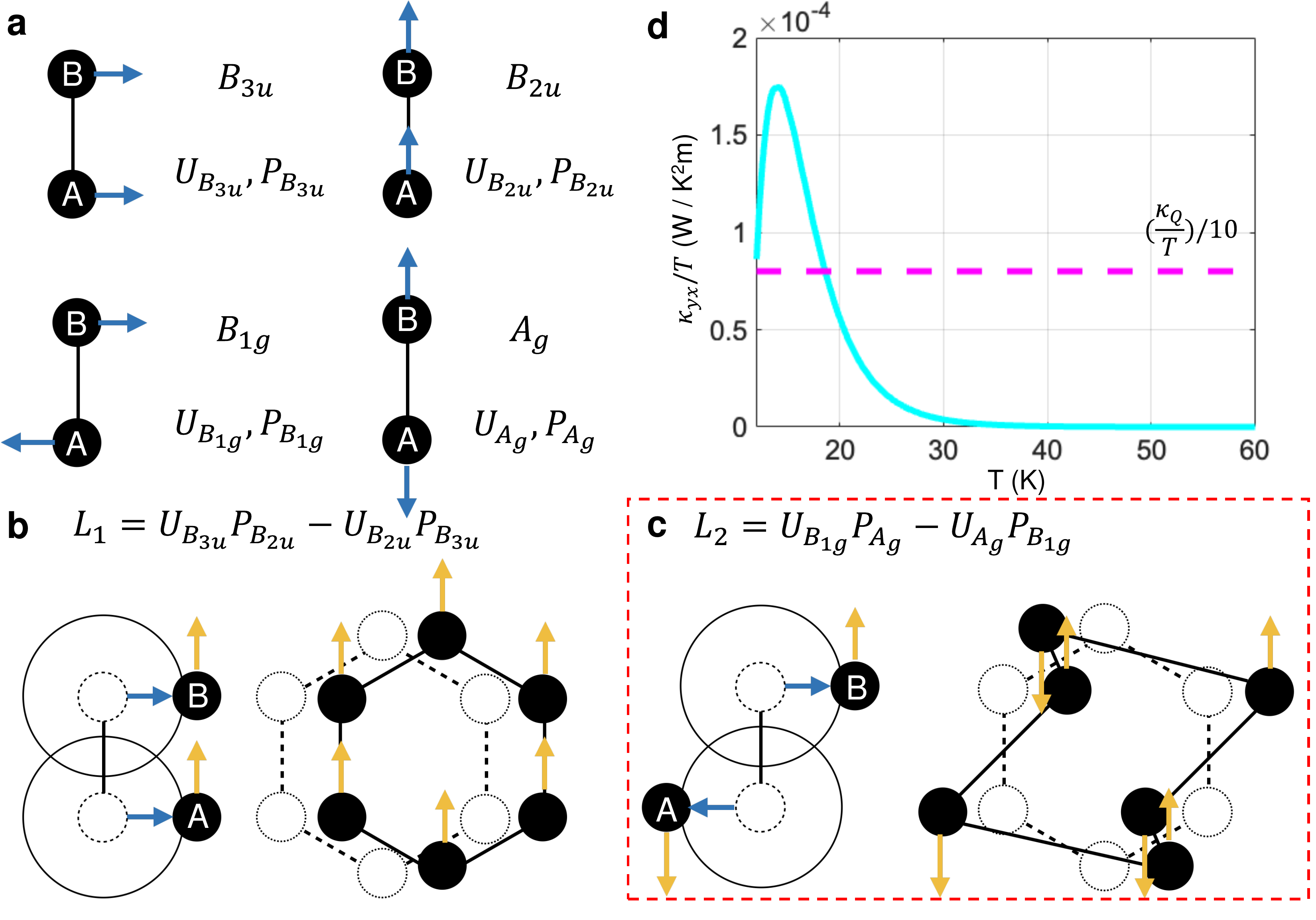}
    \caption{{\bf The chiral phonon coupling to SSC and associated THC.}
    (a) The IRREP classfication of nuclei displacement and momentum in the $xy$ plane. Here, the arrows represent either displacement and momentum. For IRREP $\mathcal{R}$, we denote the displacement $U_{\mathcal{R}}$ and momentum $P_{\mathcal{R}}$, which is defined in Eq.~(\ref{eq:U}).
    The SSC $\chi$ is in $B_{1g}$. 
    (b-c) The types of chiral phonons that can couple to the SSC. Here, blue arrows are the displacement, and yellow arrows are the momentum. 
    However, for (b), since the threefold rotation symmetry is kept, the emergent field vanishes. 
    Therefore, (c) with broken threefold rotation symmetry can couple to the SSC. 
    (d) The computed THC from the acoustic phonon of honeycomb lattice based on $\A$-RuCl$_3$ parameters (solid line) and its comparison with half-quantized THC $\kappa_Q/T$ (dotted line).
    This shows the features like the peak near $15$ K and the slow decay, which was observed in the experiments.
    }
    \label{fig:4}
\end{figure}

The honeycomb lattice consists of two sublattices, $A$ and $B$, as shown in Fig.~\ref{fig:4}(a). Its space group is $P6/nmm$, and the point group of the unit cell center is $mmm$, which includes identity $1$, twofold rotations $2_x,2_y,2_z$, inversion $-1$, and the mirrors $m_x,m_y,m_z$. 
Accordingly, there are eight irreducible representations (IRREPs): $A_g$, $B_{1g}$, $B_{2g}$, $B_{3g}$, $A_u$, $B_{1u}$, $B_{2u}$, and $B_{3u}$~\cite{aroyo2006bilbao,aroyo2006bilbao2,aroyo2011crystallography}. 
Following the canonical scheme, we classify nuclei displacements and momenta in the $xy$ plane according to these IRREPs, as depicted in Fig.~\ref{fig:4}(a). 
For an IRREP $\mathcal{R}$, we denote the nuclei displacement as $U_{\mathcal{R}}$ and the nuclei momentum as $P_{\mathcal{R}}$.
Explicitly, the classified nuclei displacements are represented by
\alg{
&U_{B_{3u}} = U_{A,x} + U_{B,x}, ~U_{B_{2u}} = U_{A,y} + U_{B,y},\n
&U_{B_{1g}} = - U_{A,x} + U_{B,x},~
U_{A_g} = - U_{A,y} + U_{B,y}. \label{eq:U}
}
Here, $\vec{U}_{A} = (U_{A,x},U_{A,y})$ and $\vec{U}_B = (U_{B,x},U_{B,y})$ represent the displacements of each sublattice. 
The nuclear momenta follow the same classification, but with $U$ replaced by $P$.
Additionally, we find that the SSC $\chi$ in a hexagon transforms as the $B_{1g}$ IRREP. 
Here, the spin configuration refers to the directional alignment of spins.
Notably, in the absence of spin-orbit coupling, no other spin configuration can couple to the phonons. 
This is because both phonons and SSC remain invariant under global $SO(3)$ spin rotations, whereas other spin configurations transform nontrivially~\cite{suzuki2017cluster,oh2018magnetic,suzuki2019multipole,oh2023transverse,oh2024phonon}. 
Furthermore, another composite order parameter $\al{\vec{S}_i \cdot \vec{S}_j}$ is invariant under $SO(3)$ and can couple to phonons. 
As $\al{\vec{S}_i \cdot \vec{S}_j}$ also changes in $B$, and can change the thermal Hall effect, it should be taken into account in principle. However, we treat it implicitly because (i) it contributes to the relaxation time which does not show a huge magnetic field dependence, and (ii) its contribution is expected to be small by the factor of $(\mu_B B/J)^2$, as it is even in the magnetic field and its leading order is $B^2$.

Two possible chiral phonons can couple to the SSC: $L_1 = U_{B_{3u}}P_{B_{2u}} - U_{B_{2u}}P_{B_{3u}}$, and $L_2 = U_{B_{1g}}P_{A_g} - U_{A_g} P_{B_{1g}}$. 
The phonon modes $L_1$ and $L_2$ are depicted in Figs.~\ref{fig:4}(b) and 4(c), respectively. 
However, $L_1$ keeps the threefold rotation symmetry $3_z$, which causes the emergent field to vanish. 
Therefore, only $L_2$~\cite{metavitsiadis2022optical} can couple to the emergent field from the SSC by breaking $3_z$. 
Using Eq.~(\ref{eq:U}), we derive the Berry connection in the form
\alg{
\vec{a}_A = \gm \chi \hat z \times (\vec{U}_A - \vec{U}_B), \vec{a}_B = \gm \chi \hat z \times (\vec{U}_B - \vec{U}_A). \label{eq:A2}
}
Here, $\gm$ is a proportionality constant. 
Therefore, the emergent field is proportional to $\vec{B}_e \propto \chi \hat z$ and the resulting form of Raman-type interaction becomes $\vec{P}_A \cdot \vec{a}_A + \vec{P}_B \cdot \vec{a}_B$.

\section{\label{sec:5} The thermal Hall conductivity}

With the emergent field established above, we now compute the THC from phonons in the honeycomb lattice. 

To compute the THC, we employ the semiclassical Boltzmann transport scheme. 
Let $\ep_{l}$ represent the phonon energy in band $n$ at momentum $\vec{k}$, where $l = (n,k)$. 
The equilibrium phonon distribution is given by the Bose-Einstein distribution $f_l^0$, and the nonequilibrium distribution is $f_l = f_l^0 + g_l$, with $g_l$ indicating the deviation from equilibrium.
Using the relaxation time approximation, the Boltzmann equation simplifies to
\alg{
\f{\partial \ep_l}{\partial \vec{k}}\cdot \grad_r T \f{\partial f_l^0}{\partial T} = - \f{g_l}{\tau_0} - \sum_{l'}w_{ll'} g_{l'.}
}
Here, $\tau_0 \sim 10^{-12}$ s is the phonon relaxation time, and $\w{ll'}$ is the antisymmetric scattering matrix derived by the Fermi Golden Rule and Born approximation:
\alg{
\w{ll'} =& 2\pi \sum_{l''}[ \{\f{\al{V_{l'l}V_{ll''}V_{l''l'}}}{\ep_{l'}-\ep_{l''}+i\eta} + c.c.\} \nm- \{ \f{\al{V_{ll'}V_{l'l''}V_{l''l}}}{\ep_{l}-\ep_{l''}+i\eta} + c.c.\}]\dt(\ep_l-\ep_{l'}). \label{eq:9}
}
Here, $V_{ll'}$ represents the impurity scattering matrix caused by the emergent field.
From Eq.~(\ref{eq:A2}) and the Raman-type interaction derived earlier, we obtain
\alg{
V_{ll'} = - B_e (T)\sum_{ab} \Gamma_{ab} (\mathcal{L}_{ab}^z)_{ll'} e^{i(\vec{k'} - \vec{k})\cdot \vec{r}_a}, \label{eq:10}
}
where $\Gamma_{ab} = \dt_{ab} - \ma^x_{ab}$, and $\vec{\mathcal{L}}_{ab} = \vec{P}_a \times \vec{R}_b$ is the generalized phonon angular momentum. 
We assume the temperature-dependent emergent field $B_e(T) =  B_{e}(T=10\text{ K})\chi(T)/\chi_{\text{max}}$, where $\chi(T)$ is obtained from Fig.~\ref{fig:2}(c) at $B=15$ T, and $\chi_{\text{max}}$ is the maximum value of $\chi(T)$ within the temperature range. 
As Figure~\ref{fig:2}(c) shows $\chi_{\text{max}} \approx 0.1$ at $B=15$ T and $T=10$ K, and Figure~\ref{fig:3}(d) compares the emergent field strength with SSC, we estimate $B_{e}(T=10\text{ K})\approx 1.16$ meV. 
Finally, the THC is computed as
\alg{
\f{\kappa_{yx}}{T} = - \f{1}{T}\sum_l \ep_l \f{\partial \ep_l}{\partial k_y} \f{h_l}{\partial_x T}, \label{eq:11}
}
where $h_l$ is the part of $g_l$ that is second order in the relaxation time $\tau_0$. 

\section{\label{sec:6} Discussion}

We compute the thermal Hall conductivity based on $\A$-RuCl$_3$ parameters and compare it with the experimental data.
Experiments estimate phonon velocities in $\A$-RuCl$_3$ as $v_\parallel \approx 4.4$ km/s for longitudinal modes and $v_\perp \approx 2.3$ km/s for transverse modes~\cite{lebert2022acoustic,mu2022role}. 
Utilizing these values, we set the longitudinal spring constant as $K_\parallel = 167.78$ meV/$\AA^2$ and the transverse spring constant as $K_\perp = 22.37$ meV/$\AA^2$.
We consider only two acoustic phonon branches since we are dealing with the temperature range well below the Debye temperature $\Theta_D \approx 209$ K~\cite{cao2016low}.
Longitudinal and transverse acoustic modes can combine to create chiral phonons through the Raman-type interaction in Eq.~(\ref{eq:10})~\cite{park2020phonon}.

The resulting $\kappa_{yx}/T$ is shown in Fig.~\ref{fig:4}(d). Several key features emerge from our results. 
First, a peak appears near $T=15$ K. 
Second, the decay of $\kappa_{yx}/T$ is gradual, remaining finite up to $T=40$ K. 
Third, the magnitude of $\kappa_{yx}/T$ is on the order of $10^{-4}$ W/K$^2$m, approximately one tenth of the half-quantized THC $\kappa_Q/T = \pi^2 k_B^2/6ha \approx 0.83$ mW$/$K$^2$m, where $h$ is Planck's constant, $k_B$ is Boltzmann's constant, and $a$ is the lattice constant of $\A$-RuCl$_3$~\cite{yokoi2021half}. 
Intriguingly, it semi-quantitatively agrees with the experimental observations in the order of magnitude~\cite{kasahara2018unusual,hentrich2018unusual,hentrich2019large,lefranccois2022evidence}, suggesting that the proposed phonon skew-scattering mechanism via SSC plays a crucial role in the observed phonon thermal Hall effect in the Kitaev model.

Experimental measurements~\cite{kasahara2018unusual,hentrich2018unusual,hentrich2019large,lefranccois2022evidence} indicate that the THC persists up to $T = 60 \sim 80$ K, whereas our results in Fig.~\ref{fig:4}(d) show a THC signal only up to $T = 40$ K. 
The minor discrepancy can be attributed to the ruthenium ions. 
As the ruthenium atom is ionized in $\A$-RuCl$_3$, these ions can directly couple to the magnetic field~\cite{agarwalla2011phonon}, potentially providing a secondary source of THC that extends the temperature range.

In addition to the phonon THC, our model predicts the emergence of a phonon angular momentum Hall effect due to the asymmetric scattering of chiral phonons with opposite angular momentum~\cite{park2020phonon}. The phonon angular momentum accumulates at the sample edges, and the accumulation could be detected via interaction with circularly polarized light.
Direct observation of the phonon angular momentum Hall effect under a temperature gradient would experimentally validate our proposed mechanism in $\A$-RuCl$_3$.

Most crucially, we anticipate that below $T<6$ K, our mechanism does not significantly contribute to the THC. 
If the emergent field strength remains constant with temperature, $\kappa_{yx}/T\propto T^2$ at low temperatures~\cite{oh2024phonon}.
In other words, as the phonon becomes inactive at lower temperatures, $\kappa_{yx}/T$ rapidly decreases. 
This confirms that our mechanism is compatible with the HQTHE below $T < 6$ K.

In addition, our mechanism provides a notable contribution to the thermal Hall effect compared to other contributions, such as magnon Hall effect and the intrinsic phonon Hall effect.
Previously, the magnon Hall effect was found to be only effective when the spin ordering exists~\cite{kim2024thermal} since magnons do not propagate above $T_N$.
However, for $\A$-RuCl$_3$, the ordering temperature is $7$ K, which is well below our temperature regime.
Also, the intrinsic phonon Hall effect results from spin-phonon interaction via spin-orbit coupling, but this system is spin-$1/2$, which exhibits small spin-orbit coupling.
Because our mechanism does not rely on spin-orbit coupling and phonons can propagate beyond the ordering temperature, the phonons can sense the SSC, and the extrinsic thermal Hall effect emerges also beyond $T_N$.

In summary, we extend an extrinsic mechanism for the phonon thermal Hall effect from Mott insulators to the Kitaev model. In this mechanism, scalar spin chirality acts similarly to magnetic impurities in the extrinsic anomalous Hall effect. Specifically, magnetic impurities cause skew scattering of spin-up and spin-down electrons, leading to the anomalous Hall effect. Similarly, scalar spin chirality can skew-scatter two chiral phonons with opposite angular momentum, producing the phonon thermal Hall effect. By extending our analysis from a Mott insulator to the Kitaev model, we gain deeper insights into spin-phonon interactions and their influence on thermal transport in strongly correlated systems.

\begin{acknowledgments}

T.O. was supported by Basic Science Research Program through the National Research Foundation of Korea(NRF) funded by the Ministry of Education (RS-2021-NR060140).
 N.N. was supported by JSPS KAKENHI Grant Numbers 24H00197 and 24H02231, and the RIKEN TRIP initiative.

\end{acknowledgments}

\end{document}